\documentstyle[preprint,aps]{revtex}
\begin{document}
\title{Three-dimensional electronic instabilities in polymerized solid
$A$C$_{60}$}
\author{S.C. Erwin$^1$, G.V. Krishna$^2$, and E.J. Mele$^2$}
\address{$^1$Complex Systems Theory Branch, Naval Research Laboratory,
Washington DC 20375\\
$^2$Department of Physics and Laboratory for Research on the Structure
of Matter,\\University of Pennsylvania, Philadelphia, Pennsylvania 19104}
\date{July 25, 1994}
\maketitle
\begin{abstract}
The low-temperature structure of $A$C$_{60}$ ($A$=K, Rb) is an ordered
array of polymerized C$_{60}$ chains, with magnetic properties that
suggest a non-metallic ground state.  We study the paramagnetic state
of this phase using first-principles electronic-structure methods, and
examine the magnetic fluctuations around this state using a model
Hamiltonian.  The electronic and magnetic properties of even this
polymerized phase remain strongly three dimensional, and the magnetic
fluctuations favor an unusual three-dimensional antiferromagnetically
ordered structure with a semi-metallic electronic spectrum.
\end{abstract}


The family of alkali-intercalated fullerides, $A_n$C$_{60}$, has a
remarkably rich phase diagram. For large alkalis (K, Rb, Cs, and their
binary mixtures) a variety of stable phases have been identified, the
electronic and magnetic properties of which are still poorly
understood.  Conventional band theory appears to explain the
insulating $n$=0 and 6 terminal phases, and suggests a conventional
metallic normal state for the $n$=3 phase, but it incorrectly predicts
that the intermediate $n$=4 phase is also metallic. Various scenarios
have been advanced to explain this discrepancy, primarily by appealing
either to very strong intramolecular electron correlations
\cite{lof,lu} or to various possible spin- or charge-density-wave
instabilities from the paramagnetic state.  This latter point of view
has received additional impetus from the recent discovery of a
polymerized form of ($n$=1) RbC$_{60}$ and KC$_{60}$ \cite{chauvet}.
Electron spin resonance data for this phase show a phase transition at
50 K, below which the spin susceptibility decreases by an order of
magnitude, suggesting an insulating ground state.  Chauvet {\it et
al.} have proposed that the single alkali electron results in a
half-filled conduction band, and that the polymer-like crystal
structure then promotes the formation of a spin-density wave that
doubles the unit cell along the polymer chain direction and opens a
gap at the Fermi level \cite{chauvet}.

In this Letter, we use the results from density-functional
electronic-structure methods to construct a physically realistic model
Hamiltonian to describe the electronic states in the low-temperature
phase of $A$C$_{60}$.  Remarkably, we find that the electronic and
magnetic properties of even this polymerized phase remain strongly three
dimensional. Indeed, by studying the fluctuations around the
paramagnetic reference state, we identify a broken-symmetry
semi-metallic ground state with a three-dimensional magnetically
ordered structure quite different from the quasi-1D structure
previously proposed for this phase.  We note, however, that the
quasi-1D electronic character hypothesized by Chauvet {\it et al.}
might be achieved by dilation of the galleries between polymer chains,
through the use of larger intercalants.  The transition from the
three-dimensional magnetic structure discussed here to a quasi-1D
structure is quite interesting and is beginning to be explored.

$A$C$_{60}$ ($A$=K,Rb) undergoes a first-order structural phase
transition around 350 K, from a high-temperature fcc phase to a
low-temperature orthorhombic phase \cite{chauvet}.  A striking feature
of the orthorhombic phase is the unusually short nearest-neighbor
fullerene distance of 9.12~\AA\ (for both K and Rb intercalants)
\cite{chauvet}.  This intermolecular spacing implies a separation
between interatomic carbon atoms on neighboring (icosahedral) fullerenes
as small as 2.0~\AA, leading to the suggestion that adjacent
molecules are actually covalently bonded to form polymer-like chains
(C$_{60}$)$_n$ aligned along one crystallographic axis
\cite{chauvet}. This picture has been confirmed by Rietveld refinement
of x-ray powder diffraction data \cite{stephens}.

To study the electronic states in such a structure, we begin by
performing electronic-structure calculations for the low-temperature
orthorhombic phase of crystalline RbC$_{60}$, using the local-density
approximation (LDA) to density-functional theory. For the details
regarding the computational methods, see Ref. \cite{erwin1} and
references therein.  We have made one structural simplification to the
Rietveld model: by rotating every chain through 45$^\circ$ {\it in the
same sense}, the Bravais lattice is changed from simple orthorhombic
to body-centered orthorhombic (bco), and the number of formula units
per unit cell is reduced from two to one.  The resulting electronic
conduction-band structure is shown in Fig.~1.  Quite surprisingly, we
find that in the occupied part of the spectrum, the dispersion
perpendicular to the chains ($\Gamma$--$M$) is quite comparable to the
dispersion along the chains ($\Gamma$--$H$). In other words, the
amplitude for electron hopping is nearly isotropic, with the resulting
Fermi surface exhibiting very strong three-dimensional character.  Two
other features of Fig.~1 deserve comment. First, the low symmetry of
this structure admits only one-dimensional irreducible
representations, so that the three-fold degeneracy of the parent
$t_{1u}$ state (of the cubic phase) is lifted.  The magnitude of this
symmetry breaking is large on the scale of the conduction band width,
of order half a volt. Second, the lowest band, which is roughly half
filled, shows positive dispersion at $\Gamma$, so that the occupied
sector of the Brillouin zone encloses the zone center. We find that
this lower band transforms like $z$, that is, these wave functions
have odd parity under reflection through the mirror plane
normal to the chains.

Qualitatively, the comparable longitudinal and transverse dispersion
is a consequence of two distinct structural features: (1) In order to
covalently bond two adjacent C$_{60}$ molecules, the valence orbitals
on intermolecular-nearest-neighbor carbon atoms essentially
rehybridize from $sp^2$ to $sp^3$.  Consequently, the $\pi$-like
conduction band in this polymerized phase contains only a very small
admixture of orbitals on intermolecular-nearest-neighbor atoms, so
that dispersion in this direction is actually governed by the
interaction of more distant neighbors \cite{pekker}).  (2) Even with
relatively weak interchain electron hopping, dispersion across the
chains also benefits from high molecular coordination in the transverse
directions.  With a larger interchain separation, this benefit would
be largely lost and intrachain hopping would predominate, leading to a
regime whose consequences are briefly discussed below.

To study the stability of this paramagnetic reference state, we
now focus our attention on the dynamics within the doped conduction
band.  We consider an effective Hamiltonian of the form
\begin{equation}
H = \sum_{i\mu,j\nu} \left( c^\dagger_{i\mu}
T_{i\mu,j\nu} c_{j\nu} +  h.c. \right) +
U \sum_{i\mu} n_{i \mu \uparrow} n_{i \mu \downarrow},
\label{hamiltonian}
\end{equation}
where $c^\dagger_{i\mu}$ creates an electron on site $i$ with orbital
polarization $\mu=x,y,z$, and the $T_{i\mu,j\nu}$ are a set of
3$\times$3 matrices giving the amplitudes for an electron hopping
between different orbitals on different sites.  The electron-electron
interaction terms describe a residual short-range repulsive potential
which, for simplicity, we take to be diagonal in both the site and
orbital indices.  For the kinetic-energy term, we retain terms
coupling each molecule to its first 12 nearest neighbors; this closes
the first coordination shell in the parent fcc structure (the
orthorhombic distortion breaks this into five inequivalent shells). We
also include a diagonal on-site term to correctly account for the
orthorhombic crystal-field splitting of the three-fold $t_{1u}$
molecular state.  The matrix elements $T_{i\mu,j\nu}$ are computed by
direct Fourier inversion of the LDA eigenvalue spectrum; this
generates the optimal tight-binding representation of the LDA bands,
and closely reproduces the first-principles single-particle spectrum.
(A detailed description of this parametrization procedure is given in
Ref. \cite{erwin2}.)  The resulting tight-binding dispersion is shown
in Fig.~1 as dotted curves; the quality of the fit to the LDA
spectrum is clearly excellent.

We now consider a mean-field decoupling of the interaction term in
Eq.~(\ref{hamiltonian}), and study the resulting (inverse) generalized
density-wave susceptibility
\begin{equation}
\chi^{-1}_{i\mu\sigma,j\nu\sigma'} = -U\left( \delta_{ij} \delta_{\mu\nu}
\delta_{\sigma,-\sigma'} + U \Pi^0_{i\mu\sigma,j\nu\sigma'} \right),
\label{susceptibility}
\end{equation}
where
\begin{equation}
\Pi^0_{i\mu\sigma,j\nu\sigma'} = - \sum_{\alpha\beta} \langle \alpha |
n_{i\mu\sigma} | \beta \rangle \langle \beta |  n_{j\nu\sigma'} |
\alpha \rangle \frac {f_\alpha - f_\beta} {E_\alpha - E_\beta},
\label{pi}
\end{equation}
and $\alpha$ labels an eigenfunction of the single-particle piece of
Eq.~(\ref{hamiltonian}) with eigenvalue $E_\alpha$ and occupation
number $f_\alpha$.  This interaction couples fluctuations in orbital
$\mu$ on site $i$ with those in orbital $\nu$ on site $j$.
Diagonalization of the susceptibility matrix then identifies both the
wave-vector and orbital character of the dominant spin- and
charge-density fluctuations in the model.  We find that for a
short-range repulsive potential, $U>0$, the spin fluctuations
within the $z$-polarized orbital are dominant. These fluctuations can
be studied further by considering the matrix
\begin{eqnarray}
 K^{zz}_{\mu\nu}(q) & = &
-\sum_{m n  k} \langle m , k | P_{\mu} \sigma _z(-q)
| n, k+q \rangle  \nonumber \\
&&\times  \langle n , k+q | \sigma _z(q) P_{\nu} |
m, k \rangle
\frac{ f_{m,k}-f_{n,k+q} } {E_{m,k} - E_{n,k+q}},
\label{piq}
\end{eqnarray}
where $P_\mu$ projects the $\mu$-orbital polarization.  By examining
the momentum dependence of the largest eigenvalue, $\kappa_{z}(q)$,
of the matrix $K^{zz}_{\mu\nu}(q)$
we find that the dominant spin-density fluctuations
are quite strongly peaked
near the $X$ point of the bco Brillouin zone. (Note that the
$\Gamma$--$X$ direction is perpendicular to the chain direction, and
the $X$ point is the center of the zone face.)  To illustrate, Fig.~2(a)
shows the static spin structure factor
\begin{equation}
S_{zz} (q) =  \frac{\kappa_{z}(q)}{1 -
(U/2) \kappa_{z}(q)},
\label{structfact}
\end{equation}
calculated at $T$=0 and $U$=250 meV.  This interaction strength is
less than the zero-temperature critical value, $U_c$=265 meV, so that
here the paramagnetic phase is stable at low temperature.  The data
are plotted in the $q_x$--$q_z$ plane, and the peaks of the
correlation function are centered at the $X$ point and its translates.
Fig.~2(b) shows the same data as a linear plot along the $\Gamma$--$X$
direction, calculated at three temperatures approaching the mean-field
ordering temperature, $T_c$, for $U$=280 meV.  Within the mean-field
theory, we find that a spin-density-wave transition at 100 K would
require a repulsive potential $U\approx 270$ meV, well within most current
estimates of the effective intramolecular repulsion given by
microscopic theories \cite{pederson,satpathy}.

A condensation at the $X$ point for $T < T_c$ describes the ordering
of a three-dimensional antiferromagnetic state with two sublattices,
such that the spin polarization alternates between the corner and
body-centered sites of the bco structure.  This spin configuration can
thus be described as ferromagnetic within each chain and
antiferromagnetic between nearest-neighbor chains.  We note that the
three-dimensional character of this magnetic state might have been
anticipated in view of the substantial three-dimensional character of
the single-particle spectrum displayed in Fig.~1.  If one views the
unpaired spins on the fullerene molecules as localized and
antiferromagnetically coupled via $J_1 > 0$ to nearest neighbors
along a cell diagonal, and antiferromagnetically via $J_2 > 0$ along
nearest neighbors within the chains, then the transition to
one-dimensional behavior requires $J_2/J_1 > 2$.  This limit is
apparently well outside the regime that actually describes these
materials.  In fact, we find that the effective intersite interaction
mediated by the polarization of the conduction electrons gives $J_2 <
0$; that is, the nearest-neighbor intrachain interaction is actually
ferromagnetic for this structure.

Fig.~3 shows the spectral density obtained at $T$=0 in the equilibrium
phase of this model; panel (a) is the density of states calculated for
the unbroken paramagnetic phase, while panel (b) is for the
equilibrium magnetically ordered phase. We find that for $U$=300 meV
the Fermi surface of the parent phase is partially gapped in the
ground-state structure, and that a well developed pseudogap is clearly
evident in the density of states. Thus the system remains semi-metallic
at low temperatures.  In the spectrum one can nevertheless clearly
observe the signature of the ferromagnetic intrachain ordering, which
essentially results in a rigid spin splitting of the lowest
($z$-polarized) conduction band, evident in the structure of the
density of states around the Fermi level.  The upper two unoccupied branches
of the conduction manifold are relatively insensitive to this ordering
transition.  We find that the pseudogap is a robust feature of this
ground state, and survives up to relatively large coupling strengths;
a fully developed gap first opens in the single particle spectrum for
$U$=470 meV.  We suggest that the decrease in the spin susceptibility
measured by Chauvet {\it et al.} \cite{chauvet} may be associated with
the opening of the pseudogap at $T_c$.  Further measurements at
temperatures much lower than $T_c$ will be required to distinguish
between insulating and semi-metallic behavior in this phase.

Our density-functional calculations indicate that the transverse
interchain hopping amplitudes in the Hamiltonian of
Eq.~(\ref{hamiltonian}) are very sensitive to the interchain
separation.  Indeed, we find that a 10\% dilation of the interchain
spacing leads to a regime where the electronic structure is
dominated by one-dimensional dynamics along the chain direction.  Here
the spin fluctuations are again dominantly antiferromagnetic in
character, but now with the ordering wave vector oriented along the
chain.  We emphasize that while this limit is apparently not realized
in K- or Rb-doped samples, it might well be achieved by doping with
substantially larger species. Experimentally, such a situation could
conceivably be realized by alloying the alkali with ``spacer
molecules'' such as NH$_3$, as has been done for $A_3$C$_{60}$
\cite{rosseinsky}. If so, the resulting magnetic phase diagram as a function
of alloying composition is expected to be quite complex \cite{mele}.

The results reported above have been obtained for a model structure
with a common orientation on each of the bco Bravais-lattice sites,
and therefore it is natural to ask how sensitive the results are to
this assumption about the orientational order.  While detailed
calculations are continuing, we should note that for the Pmnn
structure (determined by Rietveld refinement) with two inequivalent
orientations per orthorhombic unit cell, our LDA calculations show
that the low-energy spectrum retains its strong three-dimensional
character \cite{mele}. Since this feature controls the physics
discussed above, we expect the scenario developed above for the
single-orientation structure still to apply.  Also, we should note
that the possibility of a quenched orientationally disordered phase
for $A$C$_{60}$ has not yet been excluded experimentally
\cite{stephens}.

In summary, we have studied the paramagnetic phase of the polymerized
fulleride $A$C$_{60}$ and have examined magnetic fluctuations around
this reference state.  We find that repulsive interactions favor a
two-sublattice three-dimensional magnetic structure for this system,
which can be stabilized at relatively modest values of the repulsion
strength. The three-dimensional character of this phase derives
primarily from the rehybridization of the cubic $t_{1u}$ conduction
band in this polymerized phase, and from the large coordination
transverse to the polymer chain axis.  Finally, we observe that
a physically plausible modification of the crystal structure, possibly
by intercalation with larger species, may be sufficient to stabilize the
quasi-1D magnetic state which had been earlier hypothesized for this
phase.

This work was supported in part by the Laboratory for Research
on the Structure of Matter (University of Pennsylvania), by the
NSF under the MRL program (Grant 92 20668) and by the DOE (Grant 91ER
45118).  Computations were carried out at the Cornell Theory
Center, which receives major funding from NSF and New York State.

%
%

%
%
%
\begin{figure}
\caption{Theoretical LDA conduction band structure (solid curves) and
tight-binding fit to the LDA spectrum (dotted curves) for polymerized
Rb$_1$C$_{60}$, in the body-centered-orthorhombic structure described
in the text.  The labeling of symmetry points is taken from the more
familiar body-centered tetragonal zone: $H$ is the zone edge along the
chain direction, and $M$ is the zone edge in the perpendicular
direction.  The Fermi level is the energy zero.
\label{bands}}
\end{figure}
\begin{figure}
\caption{(a) Surface plot of the static spin structure factor,
$S_{zz}(q)$, for $T$=0 and $U$=250 meV, plotted in the $q_x-q_z$
plane. The sharp peak at the $X$ point of the bco zone corresponds to
the three-dimensional antiferromagnetic ordering described in the
text. (b) $S_{zz}(q)$ plotted along $\Gamma$--$X$, for $U$=280 meV and
shown for three temperatures approaching the ordering value, $T_c$=130
K.
\label{spinstructure}}
\end{figure}
\begin{figure}
\caption{Spectral density at $T$=0 for (a) the paramagnetic
reference state, and (b) the magnetically ordered ground state.
The Fermi level is the energy zero.
\label{dos}}
\end{figure}
\end{document}